\documentclass[aps,preprintnumbers, amsmath, showpacs,]{revtex4}

\usepackage{epsfig}

\usepackage{graphicx}
\usepackage{dcolumn}
\usepackage{bm}

\usepackage{israphy}                                                    

\begin{document}

\title{ BCS-BEC Crossover and Stability in a Nambu-Jona-Lasinio Model with Diquark-Diquark Repulsion}

\author{Efrain J. Ferrer, Vivian de la Incera, Jason P. Keith and Israel Portillo}
\affiliation{Department of Physics, University of Texas at El Paso, 500 W. University Ave., El Paso, TX 79968, USA}



\date{\today}


\begin{abstract}
We investigate the equation of state (EoS) along the BCS-BEC crossover for a quark system described by a Nambu-Jona-Lasinio model with multi-fermion interactions. Together with attractive channels for particle-antiparticle ($\Gs$) and particle-particle ($\Gd$) interactions, a multi-fermion channel  with coupling $\lambda$ that accounts for the diquark-diquark repulsion is also considered.  The chiral and diquark condensates are found in the mean-field approximation for different values of the coupling constants.  The parameter values where the BCS-BEC crossover can take place are found, and the EoS is used to identify the stability region where the BEC regime has positive pressure. We discuss how the particle density and the repulsive diquark-diquark interaction affect the stability window in the $\Gs-\Gd$ plane and find the profile of the pressure versus the density for various values of $\lambda$ and $\Gd$. The effects of $\lambda$ and $\Gd$ in the BCS-BEC crossover tend to compensate each other, allowing for a feasible region of densities where the crossover can occur with positive pressure. These results, although mainly qualitative, should serve as a preliminary step in the microscopic analysis required to determine the feasibility of the BCS-BEC crossover and its realization in more realistic models of dense QCD that can be relevant for applications to neutron stars. 
    
\end{abstract}

\pacs{12.38.Mh, 03.75.Nt, 26.60.Kp, 24.85.+p}

\keywords{QCD phases, high dense quark matter, BCS-BEC crossover}
\maketitle


\section{Introduction}    
In the last decade, much effort has been devoted to mapping the phases and understanding the properties of quantum chromodynamics (QCD) in extreme conditions of temperature and/or density \cite{QCDphases}. The QCD phases at high temperatures or densities are readily described in terms of quark and gluon degrees of freedom thanks to the mechanism of asymptotic freedom. The quark-gluon plasma, where all the QCD symmetries are unbroken, occurs at high temperatures and low densities. Its existence has been experimentally confirmed by the Relativistic Heavy-Ion Collider (RHIC) at Brookhaven National Laboratory and the Large Hadron Collider (LHC) at CERN. At low temperature, with increasing density, the hadronic phase gives way to a degenerate Fermi system of quarks. What happens at this point is a topic of intense debate nowadays. If the density is very large, the attractive quark-quark channel embedded in QCD, makes the system unstable against the formation of Cooper pairs and a Bardeen-Cooper-Shriffer (BCS) color superconductor is formed. However, if the density is large but not large enough, the system develops chromomagnetic instabilities \cite{Chromo-Inst}. What phases realize in this region is still unknown, although there is a broadly accepted expectation that they should be inhomogeneous, as reflected in several propositions involving momentum-dependent order parameters \cite{Kaons}-\cite{Vortex}.  

At intermediate densities, other channels besides the diquark's may be also relevant and new exotic phases with quark-hole pairing could also produce a different kind of inhomogeneous ground state \cite{DW}. 
A much simpler possibility, also considered in the literature \cite{Strong-Coupling-1, Strong-Coupling-2}, is to avoid the chromomagnetic instabilities by assuming that in the region of intermediate densities the strength of the strong coupling is large enough to favor a stable realization of color superconductivity.

If there is indeed a strong coupling region of QCD where color superconductivity is stable, one may wonder if such a system would crossover from a BCS to a Bose-Einstein Condensate (BEC) state. The phenomenon of the BCS-BEC crossover has been extensively investigated in strongly-correlated fermion systems like ultracold atoms \cite{cold-atoms}, where the attractive coupling among fermions is tuned with an external magnetic field via a Feshbach resonance \cite{Feshbach}. The smooth crossover so induced drives the system from a BCS superfluid to a BEC state of composite molecules \cite{Crossover}. Even though there is no change of symmetry and hence no phase transition but just a crossover between these two regimes, their distinct features can be readily identified. In the BCS side the coherence length of the pairs is much larger than the mean inter-particle distance and consequently, the properties of the system are basically determined by fermionic degrees of freedom. In contrast, in the BEC side, the strong interaction bounds the fermions of the Cooper pair into a bosonic molecule; thus no fermionic degrees of freedom remain. During the crossover, a mix of these two tendencies is present. Despite their apparent differences, there are common features of QCD and cold atoms that suggest the possibility of a crossover from a color-superconducting BCS dynamics to a BEC one in cold and dense QCD. However, in QCD the tuning of the strong interaction triggering the crossover is due to the density and not to a Feshbach resonance. The BCS-BEC crossover in QCD has been explored by many authors using various relativistic models and different techniques \cite{Abuki}-\cite{Jason}.

The low temperature, high density region of the QCD phase diagram is still unreachable in terrestrial experiments. Luckily, the dense and relatively cold cores of neutron stars may offer a natural medium where quarks could be deconfined. Neutron stars can then be used as natural testing objects to probe the feasibility of the high-to-intermediate dense phases of quark matter that have been proposed theoretically. For this to take place, one would need to identify observable signatures that could be connected to the inner phase of the star. This in turn requires studying the corresponding equation of state (EoS), since many properties of the neutron stars are linked to their EoS.  

It is easy to understand that a star core made of a color superconducting matter crossing over to the BEC regime might not be stable. A compact star made of non-interacting bosons would collapse because at zero temperature all the bosons in the core would occupy a single state of zero momentum and would not generate the matter pressure required to compensate the gravitational pull.  This issue has been long recognized \cite{Boson-Stars-Rev}. A possible solution is to consider a repulsive interaction between the bosonic molecules that prevent them to condense into a pressureless BEC state. This idea has been discussed by several authors \cite{BS-2}, who explored the effect of the repulsion within an effective theory of bosons mimicking the diquarks.  However, these studies ignored the fact that the same channel producing the repulsion between the diquarks, can also affect the gap equation and the fermion dispersion relations, which play a very important role in the crossover, dictating if the system is in a bosonic or fermonic regime. In consequence, a self-consistent study of the crossover and the stability of the system would require to take into account the presence of the repulsive interaction at an earlier step, i.e. within the effective fermion theory considered. A first attempt in this direction was done in \cite{Jason}, using a mixed treatment since the diquark-diquark channel was introduced in the fermion Lagrangian, but it was only taken into account at the tree level. Besides, the vacuum pressure was subtracted through a bag constant B whose value was not determined self-consistently from the model parameters, so the results were strongly B-dependent.

In the present paper, we aim to investigate the realization of the BCS-BEC crossover in a simple, QCD-inspired NJL model of quarks at intermediate densities where chiral and diquark condensates can in principle coexist. The proposed NJL model contains a diquark-diquark repulsion in the form of an eight-order fermion interaction term that affect the mean-field Lagrangian and the dispersions. In addition, the vacuum contribution is determined self-consistently in terms of the model parameters. We explore whether the system can crossover to the BEC regime, under which conditions this can occur, and how these conditions affect the EoS and the stability of the BEC phase.  By stable BEC regime we mean one where the pressure is positive. A BEC regime then may exist if, for a given choice of the couplings, there is a range of densities leading to positive pressures. For applications to neutron stars, a positive matter pressure is required to equilibrate the gravitational pull of the star.  In addition, for the BEC regime to exist in the star, the range of densities required for its formation must be also physically attainable.  Any density producing negative values of the pressure is of course forbidden in the star. In addition, any density smaller than the nuclear saturation density $\rho_S$ forbids quark deconfinement and hence cannot sustain quark matter. The toy model used in our analysis will allow us to easily extract the qualitative physical behavior in the crossover region and to establish its reliability and the stability of the strongly interacting quark system. Nevertheless, this work should be interpreted as a first step to obtain valuable insight that can then be useful to explore more realistic models with better potential for quantitative predictions. 

The paper is organized as follows. In Section \ref{2}, we introduce the NJL model with multi-fermion interactions and use it to find the mean-field thermodynamic potential and the minimum equations that will be used later to find the dynamical parameters and to study the crossover. In Section \ref{3}, the crossover transition and its stability are numerically explored. The gap, the chiral condensate, and the quark chemical potential are all obtained, at fixed density, as functions of the strength of the diquark coupling $\Gd$ for various values of the diquark-diquark repulsion $\lambda$.  The EoS is then obtained in terms of $\Gd$ and for fixed values of $\Gs$ and density, and the stability window in the $\Gs-\Gd$ plane for different densities and $\lambda$ values is determined. In Section \ref{4}, the profile of the pressure versus the density is found for various values of the diquark-diquark and quark-quark couplings. We discuss the relevant physical behaviors emerging from the numerical findings and their implications for the realization of the BCS-BEC crossover in neutron stars. Section \ref{5} contains the concluding remarks of this work.
\section{Diquark and Chiral Condensates in an NJL Model with Diquark-Diquark Repulsion}\label{2}

\subsection{An NJL Model with Multi-Fermion Interactions}

   Let us consider a simple NJL model of Dirac fermions with the usual scalar fermion-antifermion interaction term and an attractive fermion-fermion interaction in the spin zero channel $J^P=0^+$,
   
 \begin{equation}\label{lagrangian}
    \mathcal{L}_1=\bar{\psi}(i\gamma^\mu \partial_\mu+\gamma^0\mu)\psi+\frac{\Gs}{4}(\bar{\psi}\psi)^2+\frac{\Gd}{4}(\bar{\psi} i\gamma_5 C \bar{\psi}^T)(\psi^T C i \gamma_5 \psi)
\end{equation}
Here $\mC=\mi\gam_0\gam_2$ is the charge conjugation matrix, $\mu$ is the baryon chemical potential, and we neglect the current fermion mass. We expect to use this model to gain insight on the phenomenon of the relativistic BCS-BEC crossover in QCD.  As known, in QCD one-gluon exchange interactions lead to a quark-antiquark channel, and also to an attractive quark-quark channel with color antitriplet, flavor antitriplet, and  antisymmetric spin  $\frac{1}{2}\bigotimes\frac{1}{2}\to0$ interactions. With a single flavor and no color degrees of freedom, the favored quark-quark channel in our simple model reduces to the spin-zero interaction channel. However, to establish a better contact with the QCD dynamics,  one should consider yet another term 
\begin{equation}\label{Int-lagrangian}
    \Lag^{dd}_{int} =\lambda\left[(\bar{\psi} i\gamma_5 C\bar{\psi}^T)(\psi^T C i\gamma_5\psi)\right]^2,
\end{equation}
which respects all the symmetries of (\ref{lagrangian}) and incorporates the repulsion between the fermion-fermion pairs, given that the same channel that favors diquark formation also gives rise to unfavorable correlations for the cross-channels that would bound quarks from the different diquarks or that would rearrange two diquarks  to form a baryon plus a single quark \cite{Wilczek}.  Introducing $\Lag^{dd}_{int}$ is equivalent to considering the repulsive $\lambda \Phi^4 $ potential used in Refs. \cite{BS-2}, where the diquarks were treated as boson fields.

The Lagrangian density of our model then becomes
  \begin{equation}\label{lagrangiantotal}
    \mathcal{L}=\mathcal{L}_1 +  \Lag^{dd}_{int}
\end{equation}
If $\lam=0$, $\mathcal{L}$ reduces to $\mathcal{L}_1$, which has been used in Refs. \cite{Abuki, He,Klimenko}.  Eight-fermion contact interaction terms in a NJL model have been previously considered \cite{Multi-Fermion} to resolve the instability of the vacuum associated to a six-fermion t'Hooft interaction. From now on, we shall refer to the fermions in our model as quarks.

Despite its simplicity,  the toy model (\ref{lagrangiantotal}) incorporates enough features of QCD to provide a qualitative picture of what should be expected when other quark's degrees of freedom besides the spin are considered. This is so because, as it will become clear throughout the paper, the essence of the BCS-BEC phenomenon is uniquely related to the change in nature of the spectrum of the quasiparticles, and this is determined by the variation of the diquark binding energy and the chemical potential with the strengths of the couplings $\Gd$ and $\Gs$. 

\subsection{Thermodynamic Potential in the Mean-Field Approximation}

To find the thermodynamic potential in the mean-field approximation, we consider chiral and diquark condensates
\begin{equation}\label{condensates}
m=-\frac{\Gs}{2}\mean{\bar{\psi}\psi},    \quad \Del=-\frac{\Gd}{2}\mean{\psi^TC\gamma_5\psi}, 
\end{equation}
and use the Hubbard-Stratonovich method to bosonize the action. After transforming to momentum space, the mean-field Lagrangian can be written as

 \begin{equation}\label{MFlagrangian}
  \mathcal{L}_{MF}= \bPsi G^{-1}(k) \Psi -\frac{m^2}{\Gs}-\chi'\frac{|\Delta|^2}{\Gd}
  \end{equation}
 in terms of the Nambu-Gorkov bispinors
\begin{eqnarray}\label{NGspinor}
\Psi=\frac{1}{\sqrt{2}}\left[\begin{array}{c}\psi \\\psi_c\end{array}\right]
\end{eqnarray}
with $\cpsi=\mC\btpsi$ the charge conjugate field, and the mean-field propagator 
  \begin{eqnarray}\label{MFPropagator}
G^{-1}(k)= \left[\begin{array}{cc} \gam^\mu k_\mu +\mu\gam^0 -m& \chi \Del^+ \\\chi \Del^- & \gam^\mu k_\mu -\mu\gam^0 -m\end{array}\right].
\end{eqnarray} 
In  (\ref{MFlagrangian}) we used the short-hand notation
\begin{equation}\label{Coeff}
    \chi=(1+32\lambda|\Delta|^2/\,\Gd^3),  \qquad \chi'=(1+48\lambda |\Delta|^2/\,\Gd^3),
\end{equation}
and introduced the gap matrices
\begin{equation}\label{Gaps}
   \Del^+= \Del \gam_5 \qquad  \Del^- = \Del^{\dag}\gam^0 \gam_5 \gam^0
\end{equation}

The corresponding mean-field partition function 
\begin{equation}\label{MFZ}
Z=\int D[\overline{\Psi}]D[\Psi]\exp i \left(\int \frac{d^4k}{(2\pi)^4}\bPsi G^{-1}(k) \Psi -\frac{m^2}{\Gs}-\chi'\frac{|\Delta|^2}{\Gd} \right)
\end{equation}
is then quadratic in the Nambu-Gorkov fields, which can be readily integrated. Going to the finite temperature formalism, the mean-field thermodynamic potential at temperature  $T=1/\beta$ and chemical potential $\mu$ takes the form
\begin{equation}\label{MF-Omega}
    \Omega=-\frac{1}{\beta V}\ln \mZ = -\frac{1}{2\bet}\sum_{n=-\infty}^\infty\int\frac{\dif^3k}{(2\pi)^3}\,\Tr\,\ln\, [\bet\mGinv(\mi\ome_n,\textbf{k})] + \chi'\frac{|\Delta|^2}{\Gd}+ \frac{m^2}{\Gs}.
\end{equation}

Doing the Matsubara sum and taking the zero temperature limit we obtain
\begin{equation}\label{Fer:Om0}
    \Omega_0 = -\lsum_{e=\pm 1}\int_\Lambda \frac{d^3 k}{(2\pi)^3}\eps_k^e + \chi'\frac{|\Delta|^2}{\Gd}+\frac{m^2}{\Gs}
\end{equation}
with $\Lam$ an appropriate cutoff to regularize the momentum integral in the ultraviolet, and $\eps_{k}^{e}$ the quasiparticle energy spectrum given by
\begin{equation}\label{Fer:def1}
    \eps_k^e = \sqrt{(\eps_k-e\mu)^2+\chi^2|\Del|^{2}}, \qquad \eps_k = \sqrt{k^2+m^2}, \qquad e = \pm 1\;.
\end{equation}
    The $e=\pm$ values denote particle ($e=+1$) and antiparticle ($e=-1$) contributions. 

\subsection{Gap Equations at Fixed Particle Number Density}

  Minimizing the thermodynamic potential (\ref{Fer:Om0}) with respect to $\Delta$ and $m$, we find respectively
\begin{equation}\label{Fer:gapEq}
    \Aint\left( \frac{1}{\eps_k^+}+\frac{1}{\eps_k^-} \right) = \frac{2(2\chi'-1)}{\Gd(3\chi^2-2\chi)}
\end{equation}
and
\begin{equation}\label{Fer:massEq}
    \Aint \frac{1}{2\eps_k} \left( \frac{\xi_k^+}{\eps_k^+}+\frac{\xi_k^-}{\eps_k^-} \right) = \frac{1}{\Gs}
\end{equation}
 with
\begin{equation}
    \xi_k^\pm = \eps_k\mp\mu
\end{equation}
As usual, to explore the BCS-BEC crossover, we consider a canonical ensemble and thus assume a fixed particle number density, $\nF=-(\pdif\Ome/\pdif\mu)$, which can be expressed in terms of the Fermi momentum $\PF$ as 
    \begin{equation}\label{Fer:denEq}
   \nF= \frac{\PF^3}{3\pi^2} = -\Aint\left( \frac{\xi_k^+}{\eps_k^+}-\frac{\xi_k^-}{\eps_k^-} \right)
\end{equation}
Eqs. \Eq{Fer:gapEq}, \Eq{Fer:massEq}, and \Eq{Fer:denEq} can now be solved numerically to find the gap, $\Delta$, chemical potential, $\mu$, and mass, $m$, as functions of the couplings $\Gd$, $\Gs$ and $\lam$.

\subsection{Equation of State}
  The energy density and pressure of the system in the zero-temperature limit are respectively calculated from
\begin{equation}\label{eq:ep}
    \varepsilon=\Omega_{0}+\mu\nF - \Omega_{vac}, ~~~~ p=-\Omega_{0} + \Omega_{vac},
\end{equation}
where
\begin{equation}\label{Fer:Om0vac2}
    \Omega_{vac} = \Omega_{0}\,(\mu\!=\!0,\Delta\!=\!0) = -\Aint\,2\,\bar{\eps}_k + \frac{m^2_{vac}}{\Gs}
\end{equation}
    with
\begin{equation}\label{Fer:hateps}
    \bar{\eps}_k=\sqrt{k^2+m_{vac}^2}
\end{equation}
 and  $m_{vac}$ determined from $(\pdif\Omega_{vac}/\pdif m_{vac})=0$.

The vacuum contribution  $\Omega_{vac}$ is added to make the pressure of the vacuum zero. In this way, we avoid the introduction by hand of a  bag constant $B$  \cite{Jason}, whose value is unknown for the system under consideration. In contrast, the vacuum contribution $\Omega_{vac}$  is dynamically determined as a function of the model parameters.

\section{Exploring the BCS-BEC Crossover and its Stability}\label{3}

\subsection{Implications of the Diquark-Diquark Repulsion for the Crossover}

The realization of the BCS-BEC crossover in the present model depends on the answers to two important questions: Is there a feasible region of parameters where the crossover can occur? Is the BEC phase stable in that region?  These questions are particularly relevant for any potential application to neutron stars. If the star's core can be qualitatively described by a quark model crossing over to the BEC phase, the connection between the BEC stability and the star's stability is fundamental. The crossover happens when the nature of the quasiparticle excitations in the presence of the diquark condensate starts to change from fermionic to bosonic. Once all the quasiparticle modes become bosonic the system is in the BEC regime, which for the present model means that no fermionic degrees of freedom remain. The quasiparticle spectrum also affects the EoS of the system and one naturally wonders if the pressure in the pure BEC region can still be positive when all the diquarks behave as bosonic molecules. It is at this point where the new ingredient, the diquark-diquark repulsion with coupling strength $\lambda$ becomes important, and we shall investigate how it may affect the region of stability.  Below, we will numerically explore the answers to all these questions.

 As a preparation for our numerical calculations, we normalize all the variables with the cutoff parameter $\Lam$ to make them adimensional. Our free parameters then are
 the Fermi momentum $\PF$, the quark-quark coupling $\Gd$, the quark-antiquark coupling $\Gs$, and the diquark-diquark repulsion $\lam$.  Adjusting them will allow us to explore the crossover region. From now on,  normalized units
\begin{equation}\nonumber
    \tGd\rarr \Gd\Lam^2, \qquad \tGs\rarr \Gs\Lam^2, \qquad  \tPF\rarr\frac{\PF}{\Lam},\qquad  \tlam \rarr \lam\Lam^8,
\end{equation}
\begin{equation}\label{Normalization}
     \tmu\rarr\frac{\mu}{\Lam}, \qquad \tDel\rarr\frac{\Del}{\Lam}, \qquad \tm\rarr\frac{m}{\Lam},\qquad  \tvep\rarr\frac{\vep}{\Lam},\qquad \tk\rarr\frac{k}{\Lam},
\end{equation}
    are used. Therefore, the cutoff parameter will not explicitly appear in the results. To understand the system dynamics, we analyze the consequence of varying the coupling constants $\tGd$, $\tGs$, and $\tlam$ for a fixed Fermi momentum $\tPF$. 

\begin{figure}[H]
    \begin{center}\resizebox{9cm}{!}
        {\includegraphics{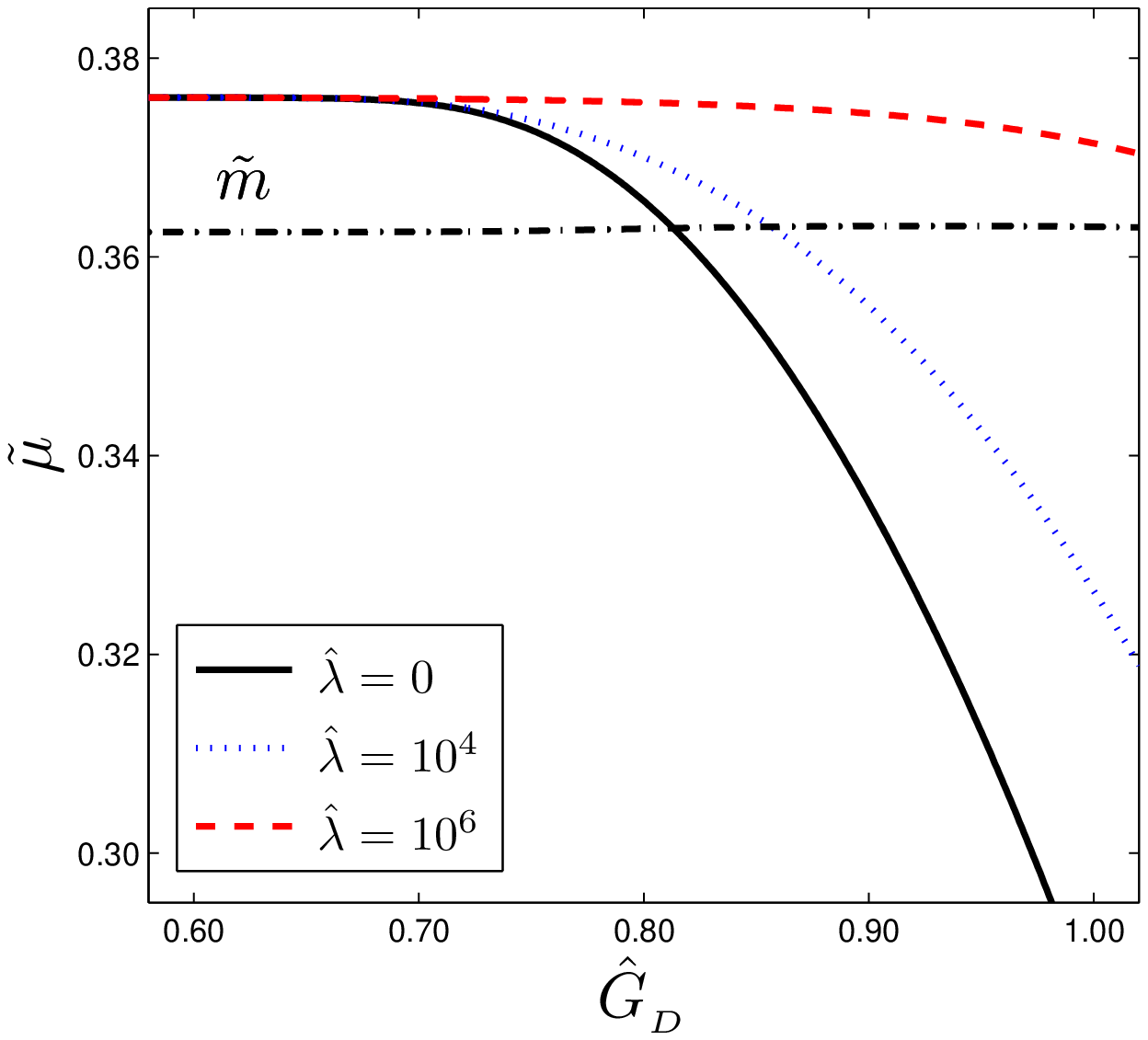}}    \end{center}
    \caption[$\tmu$ and $\tP$ Vs $\hGd$ at $\tPF=0.10$ and $\hGs=1.20$ ]{Behavior of the chemical potential $\tmu$ and the dynamical mass $\tm$ with the diquark coupling $\hGd$ for different values of $\hlam$ at $\tPF=0.10$ and $\hGs=1.20$}\label{Fig:Fer:Gd_Mu}
\end{figure}

As shown in \Fig{Fig:Fer:Gd_Mu}, the fermion chemical potential decreases with the increase of the diquark interaction, while the dynamical mass remains practically unchanged. For convenience, we use redefined normalized quantities $\hat{Q} \equiv \frac{\widetilde{Q}}{4\pi^2}$ in  the graphs. The decline of $\mu$ is crucial for the onset of the crossover to the BEC phase, since a main condition for a relativistic gas to become bosonic is $\mu < m$ \cite{Haber}. The value of $\hGd$ at which $\mu=m$ is the critical $\hGdcr$ for the BCS-BEC crossover. For $\hlam=0$, we find the critical value, $\hGdcr= 0.81$, beyond which the condition $\mu < m$ holds.   It is also apparent that the diquark-diquark repulsion tends to expand the region of $\hGd$'s where the BCS regime is favored.

\begin{figure}[H]
    \begin{center}\resizebox{9cm}{!}
        {\includegraphics{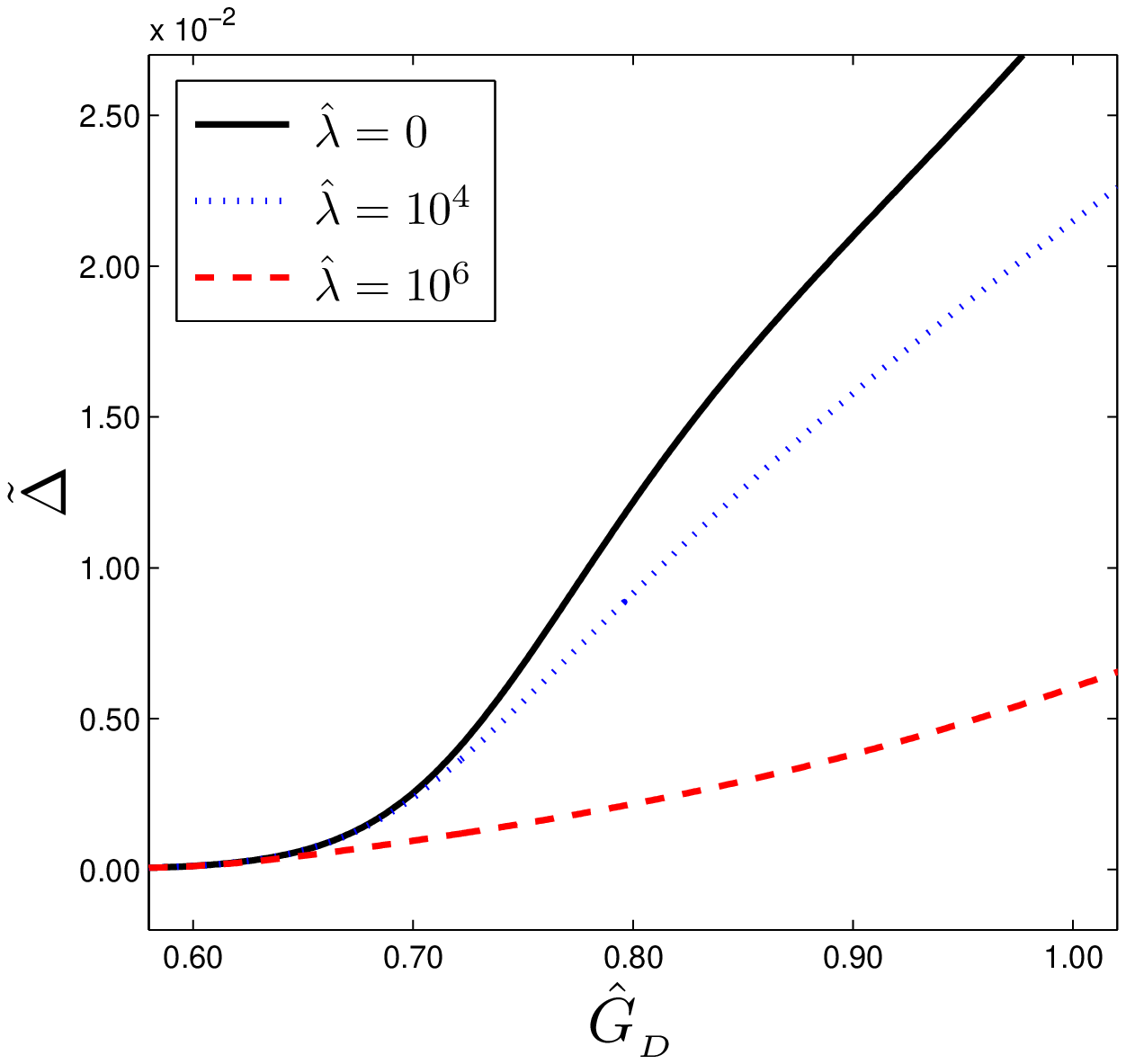}}  \end{center}
    \caption[$\tDel$ and $\tvep$ Vs $\hGd$ at $\tPF=0.10$ and $\hGs=1.20$ ]{Gap parameter $\tDel$ vs diquark coupling $\hGd$ for different values of $\hlam$ at $\tPF=0.10$ and $\hGs=1.20$}\label{Fig:Fer:Gd_Del}
\end{figure}

 In \Fig{Fig:Fer:Gd_Del} we plot the variation of the gap parameter with the diquark coupling $\hGd$. As expected,  the gap parameter increases with $\hGd$, but this effect is damped by $\hlam$. This is in agreement with the tendency of $\hlam$ to reinforce the BCS region, already seen in \Fig{Fig:Fer:Gd_Mu}.  Hence, $\hGd$ and $\hlam$ tend to create opposite effects. While the former increases the gap, or equivalently decreases the correlation length of the diquarks, making them to behave more as bosonic molecules, the repulsion between the diquarks tend to hold these molecules apart from each other, counteracting their inclination to condense into a single state. The competition between these two tendencies will ultimately determine whether the system can cross over or not to the BEC region and if it does, whether that regime can have or not a positive pressure. 

  We call attention to the fact that as the crossover critical point is approached, the ratio $\Delta/\mu$ increases. Note that the phenomenology explored here is very different from the one considered in \cite{Hen-2010}, where the diquarks were generated in a two-color, two-flavor NJL model characterized by a quantum phase transition point at $\mu=m_\pi/2$. That was a second-order phase transition at $T=0$ separating a phase with no diquarks from one with diquark condensation that breaks the $U_B(1)$ symmetry. Near such a quantum phase transition, the gap $\Delta$ is vanishingly small and a Ginzburg-Landau (GL) expansion of the mean-field free energy in powers of  $\Delta$ can be carried out. The GL free-energy in the two-color model then reduces to the Gross-Pitaevskii \cite{Gross-P}  free energy that describes a weakly repulsive Bose diquark system. This is not the case in the BCS-BEC crossover phenomenon discussed in the present work. The transition from BCS to BEC regime is a crossover, not a phase transition. No symmetry breaking takes place during the crossover, thus the symmetry is the same in the two regimes. The difference between these two regimes is only reflected in the character of the dispersions, being fermionic or bosonic, or equivalently, in the comparison between the coherence length of the pairs and the mean inter-particle distance. In the crossover region the free energy cannot be expanded in powers of $\Del$ because in this region this parameter is not small.  In our model the diquark-diquark repulsion has to be introduced as an extra interaction term parametrized by a new coupling constant $\lambda$.
 
\begin{figure}[H]
    \begin{center}\resizebox{14cm}{!}
        {\includegraphics{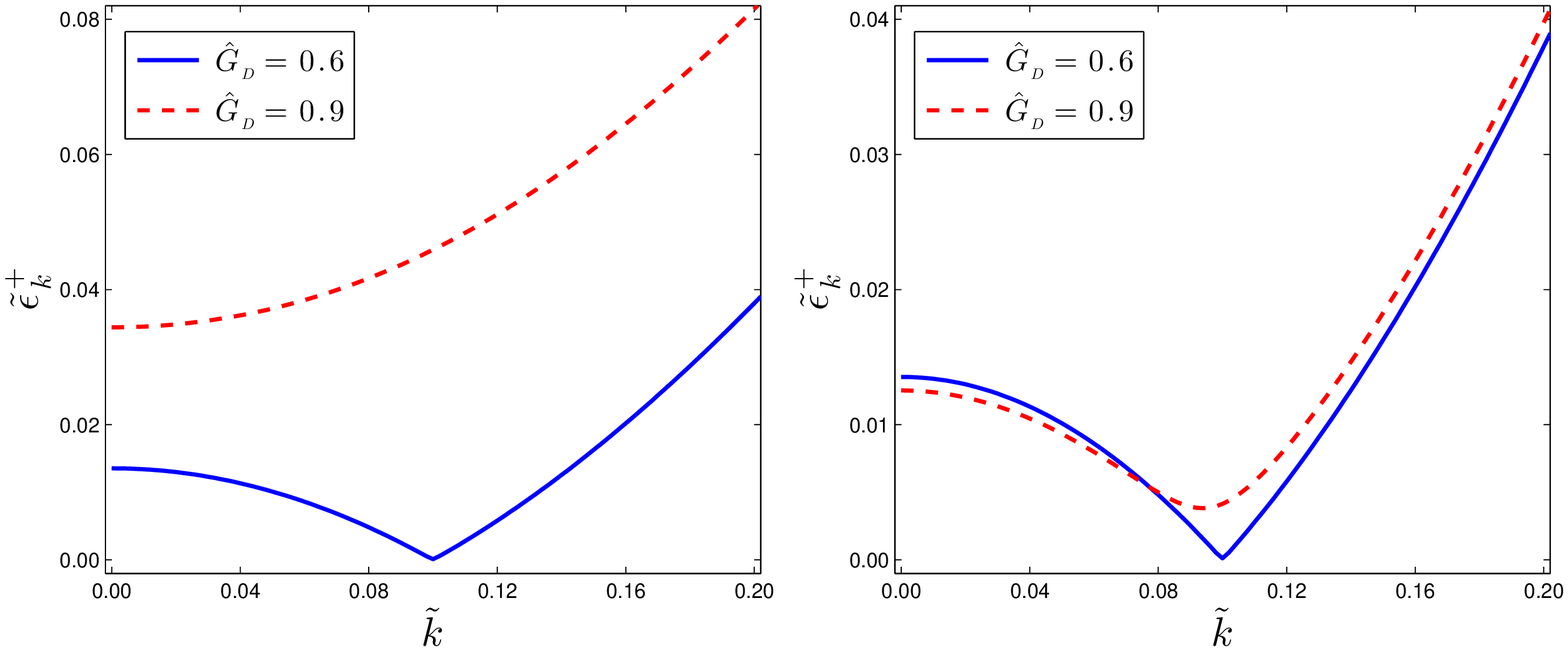}}    \end{center}
    \caption[Spectrum $\tvep_k^+$ before and after the crossover]{Effects of the couplings on the quasiparticle dispersions. Left panel has $\hlam=0$; right panel has $\hlam=10^6$.}\label{Fig:Fer:Spectrum}
\end{figure}

One can corroborate the effects of  $\hGd$ and $\hlam$ on the crossover by looking at the dispersion modes of the quasiparticles. In the left panel of \Fig{Fig:Fer:Spectrum}, we plotted the quasiparticle dispersions for $\hGd$ smaller and larger than $\hGdcr$, both at  $\hat{\lambda}=0$.   Clearly two qualitatively different type of modes appear here. For $\hGd < \hGdcr$, the minimum of the dispersion occurs at $\tk =\sqrt{\tmu^2 - \tm^2}$, with excitation energy given by the gap $\Del$, a behavior characteristic of quasiparticles in the BCS regime. In contrast,  for $\hGd>\hGdcr$, the minimum of the dispersion occurs at $\tk = 0$, with excitation energy $\epsilon^+=\sqrt{(\tmu-\tm)^2+\Del^2}$,  typical of a bosonic-like quasiparticle. Therefore, the dispersion is associated with the BCS (BEC)-regime when $\hGd < \hGdcr$ ($\hGd > \hGdcr$). For comparison, in the right panel we plotted the dispersions for the same $\hGd$'s, but with $\hat{\lambda}=10^6$. In this case the two dispersions are still fermionic-like, in concordance with the BCS regime, a behavior already noticed in \Fig{Fig:Fer:Gd_Mu}. Therefore, a large enough diquark-diquark repulsion can modify the nature of the dispersions from bosonic to fermionic.
     
\subsection{Diquark-Diquark Repulsion Effects on the EoS and Stability Region}

\begin{figure}[H]
    \begin{center}\resizebox{9cm}{!}
        {\includegraphics{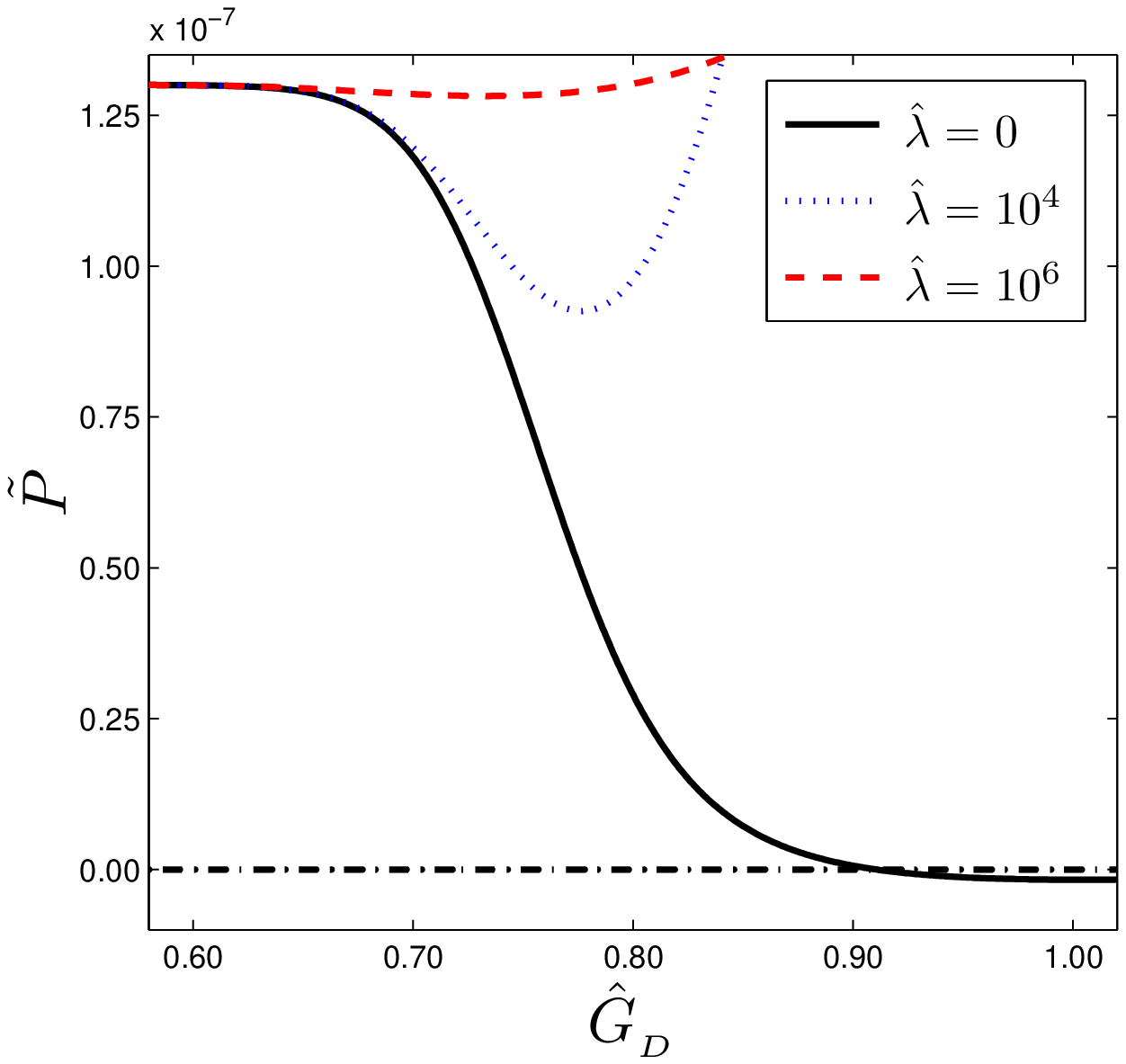}}    \end{center}
    \caption[$\tmu$ and $\tP$ Vs $\hGd$ at $\tPF=0.10$ and $\hGs=1.20$ ]{Change of pressure $\tP$ with the diquark coupling $\hGd$ for different values of $\hlam$ and at $\tPF=0.10$ and $\hGs=1.20$.}\label{Fig:Fer:Gd_Mu_Pres}
\end{figure}

Let us turn our attention now to the EoS and how it is affected by the couplings in the crossover region. In \Fig{Fig:Fer:Gd_Mu_Pres}, we have plotted the pressure versus $\hGd$ for various values of $\hlam$ at $\tPF=0.10$ and $\hGs=1.20$. When $\hlam=0$,  the system crosses over to the BEC state at $\hGdcr=0.81$, as found in \Fig{Fig:Fer:Gd_Mu}. At $\hGd\sim 0.9$ the pressure crosses the zero line, becoming negative.  Therefore, only in the region between $\hGd=0.81$ and $\hGd\sim 0.9$, the BEC regime is stable. Switching on the diquark-diquark repulsion slows down the falling of the pressure, which for $\hat{\lambda} \geq \hat{\lambda}_{cr}=11$ never vanishes.  If one keeps increasing $\hlam$ the pressure develops a nonzero minimum at $\hGd < 0.9$ that becomes shallower for even larger $\hlam$. This happens for values of $\hlam$ several orders of magnitude larger than the one required to ensure stability. The fact that with increasing diquark-diquark repulsion the window of stability expands and the crossover exists for a larger range of parameters should be model-independent, because despite the shortcomings of the simple model used in our calculations, we expect that our results encompass the main physics features that should be relevant for the possible realization of the BCS-BEC crossover in neutron stars with quark matter cores. 

The role of a diquark-diquark interaction in avoiding a pressureless state was previously discussed in \cite{BS-2}, where the system's Lagrangian was written only in terms of a boson field  $\Phi$ that represented the diquarks, and a repulsive interaction modeled with a $\lambda \Phi^4$ potential. In our case, by keeping the fermion degrees of freedom, we can self-consistently treat the diquarks as composite bosons and explore the influence of the couplings and thermodynamical parameters on the crossover region and the EoS. 

\begin{figure}[H]
    \begin{center}\resizebox{9cm}{!}
        {\includegraphics{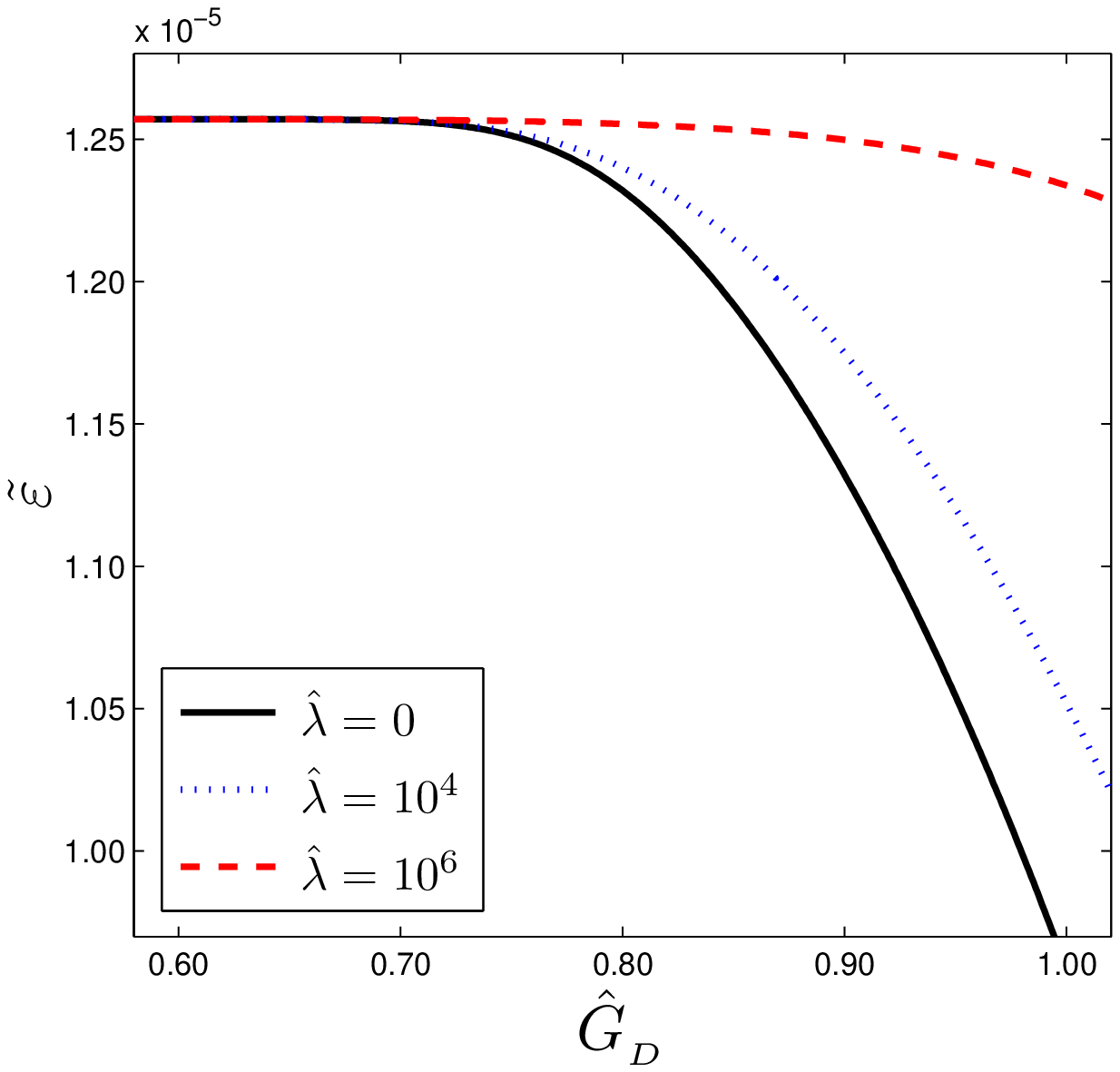}}    \end{center}
    \caption[$\tDel$ and $\tvep$ Vs $\hGd$ at $\tPF=0.10$ and $\hGs=1.20$ ]{Energy density $\tvep$  as a function of the diquark coupling $\hGd$ for different values of $\hlam$ at $\tPF=0.10$ and $\hGs=1.20$}\label{Fig:Fer:Gd_Del_Ene}
\end{figure}

The plot of the energy density $\hvep$ versus $\hGd$ in  \Fig{Fig:Fer:Gd_Del_Ene} shows a behavior consistent with the growing of the gap $\Del$ with $\hGd$ observed in  \Fig{Fig:Fer:Gd_Del}. When diquark formation is energetically favored, a larger gap decreases the energy even more, since it increases the energy used for condensation. On the other hand, the repulsion among diquarks smooth out the plunge of $\hvep$, since it decreases $\Del$ for each given value of $\hGd$. 
 
\begin{figure}[H]
    \begin{center}\resizebox{16cm}{!}
        {\includegraphics{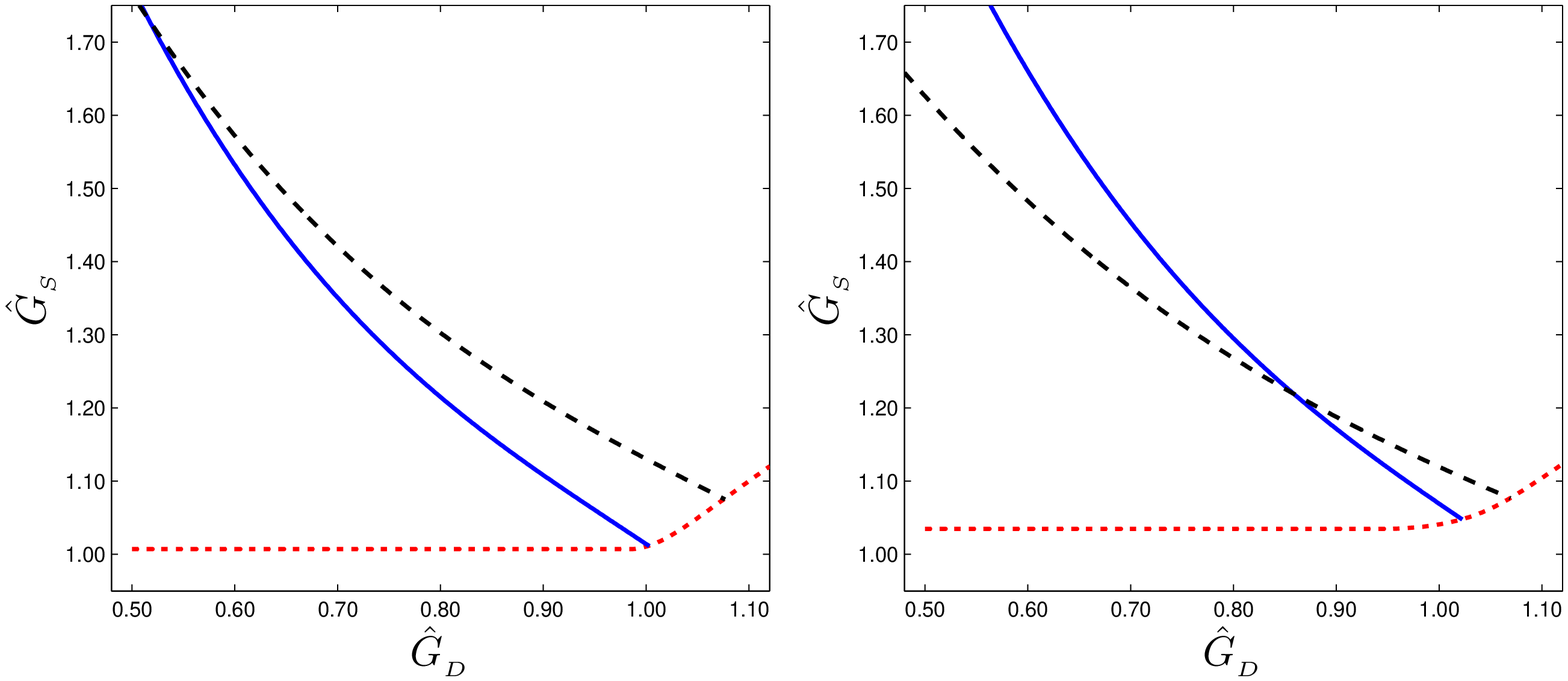}}    \end{center}
    \caption[Region on which the BSC-BEC Crossover take place]{Stability regions for the BEC state near the BCS-BEC crossover. The Fermi momentum is $\tPF=0.10$ on the left panel and $\tPF=0.20$ on the right panel. Both graphs were found for $\hlam=0$}\label{Fig:Fer:crValues1}
\end{figure}

    \Fig{Fig:Fer:crValues1} displays the boundaries in the $\hGd$-$\hGs$ plane defining the region of parameters where the crossover can occur and where the BEC state is stable. Left panel corresponds to $\tPF=0.10$ and right panel to $\tPF=0.20$. The solid line marks the crossover condition (points where $\mu=m$). To the right of this line the BEC regime develops. The dashed line denotes the zero-pressure condition, which separates a negative pressure regime to the right from a positive pressure one to the left. The short-dashed line separates the massless region (below), characterized by zero chiral condensate, from the massive region (above). The stability window, defined as the region with positive pressure where the BEC regime can take place, is the one enclosed by the three lines. Comparing the two graphs we see that the stability region shrinks as $\tPF$ increases, that is, a larger density tends to favor BCS over BEC, as physically expected. For fixed density and $\hGd$, the stability window narrows for larger $\hGs$, indicating that the difference between the system and vacuum pressures becomes smaller with larger chiral coupling.  The stability window completely disappears at $\hlam=0$ when $\tPF=0.23$.   
    
\begin{figure}[H]
    \begin{center}\resizebox{16cm}{!}
        {\includegraphics{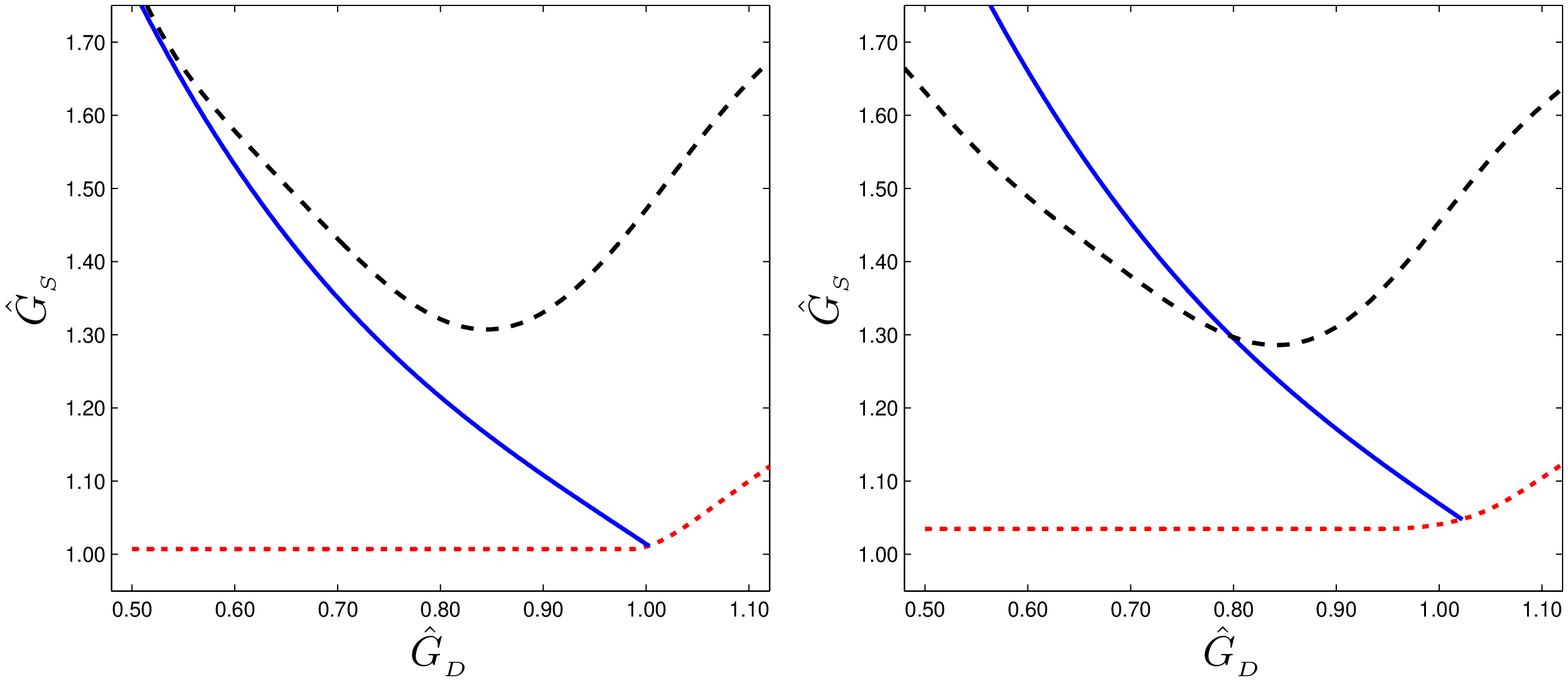}}    \end{center}
    \caption[Region on which the BSC-BEC Crossover take place]{Stability region on which the BSC-BEC crossover takes place for $\hlam=50$. On the left panel $\tPF=0.10$, and on the right panel $\tPF=0.20$}\label{Fig:Fer:crValues2}
\end{figure}

If we switch on a nonzero $\lam$, the dashed line moves up, while the solid line almost remains the same, and thus the stability region expands in the $\hGd$-$\hGs$ plane. \Fig{Fig:Fer:crValues2} shows the stability windows at $\hlam=50$ for two different Fermi momenta, $\tPF=0.10$ (on the left panel) and $\tPF=0.20$ (on the right panel). The  stability window still shrinks with a larger particle density, but it now covers a larger parameter space than at $\hlam=0$. Note that despite the simplicity of the model considered in this analysis, there is always a region of stability for $\hGd< \hGs$.  As known \cite {Wilczek}, the one-gluon exchange interaction leads to four-fermion point interactions satisfying $\Gd< \Gs$, so it is natural to expect the existence of similar windows of stability in more realistic models where $\hGd$ should be constrained to values smaller than $\hGs$ only.

These results underline the relevance of the diquark-diquark repulsion in the study of the BCS-BEC crossover and the determination of a stable BEC regime when going to more realistic models with physically viable values of the couplings and particle density. 


\section{Neutron Stars and Reliability of the BCS-BEC Crossover} \label{4}

\begin{figure}[H]
    \begin{center}\resizebox{14cm}{!}
        {\includegraphics{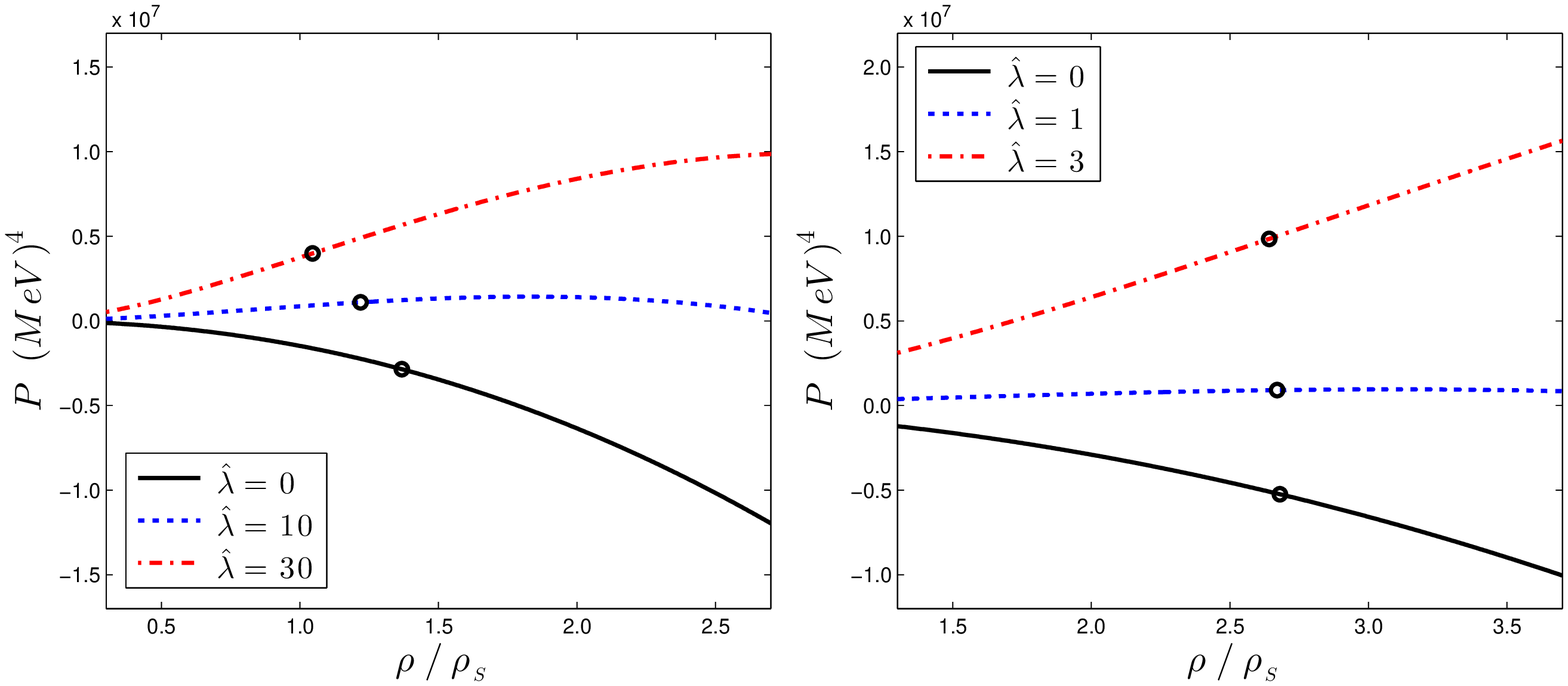}}    \end{center}
    \caption{Change of pressure $P\,(\MeV)^4$ with density in units of $\rhos$ for $\Gs/4=2.06\Lambda^{-2}$, $\Lambda=664.3 \MeV$, and for different values of $\hlam$. On the left panel $\eta=0.75$; on right panel $\eta=0.90$ }\label{Fig:Fer:Den_P}
\end{figure}

To infer the implications of our findings for neutron stars, we mimic a more realistic realization of color superconductivity within our simple model by adding three colors and two flavors, but keeping the same simple gap structure of the original toy model (\ref{lagrangian}). The profile of the pressure versus the density (given in multiples of the saturation density), when we include these extra degrees of freedom, is given in \Fig{Fig:Fer:Den_P}. The plots were done for fixed $\Gs/4=2.06\Lambda^{-2}$, with $\Lambda=664.3 \MeV$, chosen to match the physical values of the pion decay constant and quark condensate in vacuum, in agreement with  the two-flavor NJL model considered in \cite{Buballa}. The two panels correspond to two values of the diquark coupling $\Gd=\eta \Gs$, $\eta=0.75$ in the left and $\eta=0.90$ in the right panel.

The circle in the curves marks the critical density for the crossover (point where $\mu=m$). The BEC region develops to the left of the circle along the curve, as the density decreases, and the BCS to its right. In the absence of diquak-diquark repulsion ($\hlam=0$), the pressure for the density region near the crossover is negative in both panels,  forbidding the realization of compact stars with such a quark matter phase.

 Nevertheless, there exists a range of $\hlam$'s where the pressure is positive and  the crossover can take place at densities larger than $\rhos$. For $\eta=0.75$ that range is $9.3 < \hlam < 38$; while for $\eta=0.9$ the range is $0.85 < \hlam < 10^3$. For values of $\hlam$ larger than the maximum of the interval, the crossover would occur at densities smaller than $\rhos$, where quarks are not deconfined, so it makes no sense at all; for $\hlam$ smaller than the minimum of the interval, the pressure near the crossover is negative, preventing such a matter state to realize in the star. Note that the diquark-diquark repulsion has a double effect, on the one hand, it increases the pressure; on the other hand, it moves the crossover to smaller values of the density. Therefore, for fixed  $\Gd$ and  $\Gs$ couplings,  $\hlam$ tends to expand the region of densities where the matter is in the BCS regime.

\begin{figure}[H]
    \begin{center}\resizebox{8cm}{!}
        {\includegraphics{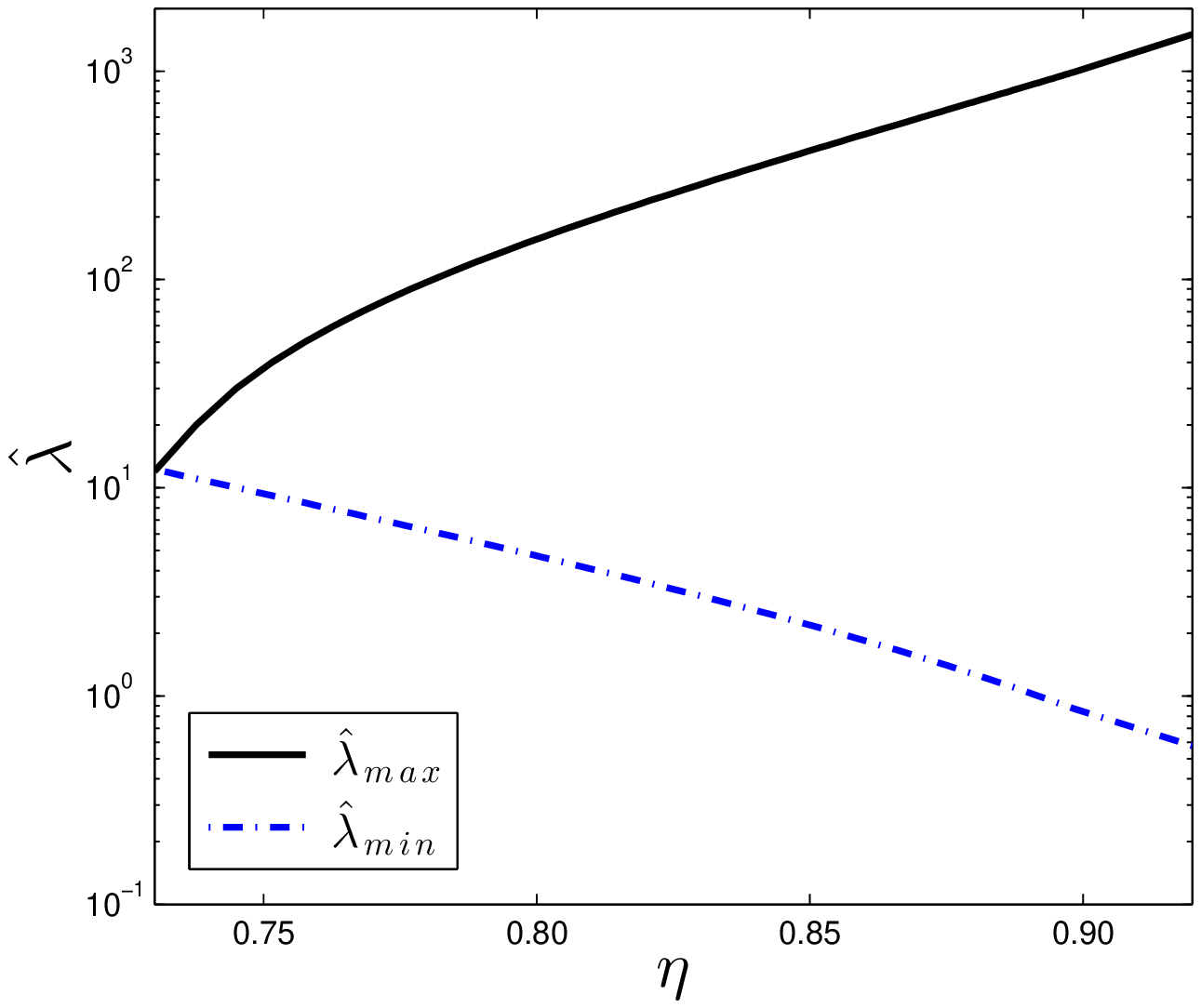}}    \end{center}
    \caption{Region of $\hat{\lambda}$ ($\hat{\lambda}_{min}<\hat{\lambda}<\hat{\lambda}_{max}$) for different $\eta$ values where the BCS-BEC crossover occurs with positive pressure and at densities $\rho > \rho_S$. }\label{Lmin-Lmax}
\end{figure}

With stronger quark-quark coupling, the crossover point moves to higher density values for each given $\hlam$, as can be gather from \Fig{Fig:Fer:Den_P},  thus the BEC regime is favored within a larger range of physically meaningful densities, in agreement with usual physical expectations. Although increasing $\hlam$ still tends to shrink the region of densities supporting the BEC regime, the effect is less pronounced with larger $\eta$. A stronger quark-quark coupling also affects the range of $\hlam$'s where the crossover is physically possible. The lower limit tends to decrease and the larger one to increase. This behavior is also apparent in \Fig{Lmin-Lmax}, which displays the parameter window --the region between the two curves-- in the $\hat{\lambda}-\eta$ plane on which the BCS-BEC crossover occurs with positive pressure and at densities larger than $\rho_S$. We found that only for $\eta \geqslant 0.73$  the crossover can be realized satisfying these conditions.

 Since the actual physical values of $\hlam$ are unknown, a larger flexibility in the values of $\hlam$'s where the crossover can occur is important for potential astrophysical applications. Despite their limitations, these results offer a strong indication that the diquark-diquark repulsion can be key to the realization of the BCS-BEC crossover in a range of densities compatible with the presence of quark matter in neutron stars. 
 
 Even though these results highlight the relevance of the diquark-diquark repulsion in the possible realization of the BCS-BEC crossover in neutron stars with a quark matter core, their predictive power is limited due to the simplicity of the model considered. Notice that such a toy model does not incorporate the color and flavor antitriplet channels responsible for Cooper pairing in more realistic descriptions of color superconductivity. Therefore, to obtain more realistic quantitative results, one would need to consider a model like the strongly coupled 2SC, with color and flavor antitriplet structures in the quark-quark interactions, extended to include the diquark-diquark repulsion term.  For neutron stars application, one would also need to impose electric and color neutrality conditions.

\section{Concluding Remarks} \label{5}

    In this paper we study the BCS-BEC crossover and the EoS for a system of quarks at finite density described by a NJL model on which diquark and chiral condensates can coexist. The model also contains an eight-fermion interaction with coupling $\lambda$, which drives the repulsion between the diquarks and ensure the stability of the system in the BEC regime for some reasonable region of the parameters. The pressure of the vacuum is not introduced as a bag constant parameter, but self-consistently determined from the vacuum free-energy. 
    
We present a detailed numerical analysis of the behavior of the crossover, the stability (positive pressure state), and the different condensates as functions of the model's parameters. The parameter values where the BCS-BEC crossover can take place are found, and the stability region where the BEC regime has a positive pressure is identified with the help of the EoS. Our findings indicate that the diquark-diquark repulsion $\lam$ tends to favor the BCS regime, so a larger $\hGd$ is needed to crossover to the BEC state. On the other hand, the stronger the coupling favoring diquark formation, the larger the region of densities where the BEC regime can exist. Hence, the effects of strong coupling $\hGd$, favoring the BEC, and strong diquark-diquark repulsion $\lam$, favoring the BCS, tend to compensate each other to allow for a feasible region of densities where the crossover can occur with positive pressure.

Our results call attention to the importance of the different channels, including the diquark-diquark repulsion, when investigating the realization of the BCS-BEC crossover tuned by the strength of the quark-quark interaction. In QCD all the coupling constants are actually density-dependent; their actual values at moderate densities will ultimately determine if a BEC regime is or not a good candidate for the phase of dense quark matter in the core of neutron stars.

    {\bf Acknowledgments:} Research reported in this paper was partially supported by the Office of Nuclear Physics of the Department of Energy under award number DE-SC0002179.


\end{document}